\documentclass[8pt,twocolumn,lettersize]{journal}

\usepackage{amsmath,amsfonts}
\usepackage{algorithmic}
\usepackage{array}
\usepackage[caption=false,font=normalsize,labelfont=sf,textfont=sf]{subfig}
\usepackage{textcomp}
\usepackage{stfloats}
\usepackage{url}
\usepackage{verbatim}
\usepackage{graphicx}
\usepackage{wrapfig}
\usepackage{balance}

\title{\textbf{Detecting the Presence of COVID-19 Vaccination Hesitancy from South African Twitter Data Using Machine Learning}}
\author{Nicholas Perikli,
Srimoy Bhattacharya, 
Blessing Ogbuokiri, 
Zahra Movahedi Nia,\\
Benjamin Lieberman, 
Nidhi Tripathi, 
Salah-Eddine Dahbi,
Finn Stevenson, \\ 
Nicola Bragazzi, 
Jude Kong, 
Bruce Mellado}
\date{}
\begin{document}
\maketitle{}
\section*{Abstract}
Very few social media studies have been done on South African user-generated content during the 
COVID-19 pandemic and even fewer using hand-labelling over automated methods.  Vaccination is a major tool in the fight against the pandemic, but vaccine hesitancy jeopardizes any public health effort. In this study, sentiment analysis on South African tweets related to vaccine hesitancy was performed, with the aim of training AI-mediated classification models and assessing their reliability in categorizing  UGC.  A dataset of 30000 tweets from South Africa were extracted and hand-labelled into one of three sentiment classes - positive, negative, neutral. The machine learning models used were LSTM, bi-LSTM, SVM, BERT-base-cased and the RoBERTa-base models, whereby their hyperparameters were carefully chosen and tuned using the WandB platform. We used two different approaches when we pre-processed our data for comparison - one was semantics-based, while the other was corpus-based. The pre-processing of the tweets in our dataset was performed using both methods, respectively. All models were found to have low F1-scores within a range of 45$\%$-55$\%$, except for BERT and RoBERTa which both achieved significantly better measures with overall F1-scores of 60$\%$ and 61$\%$, respectively. Topic modelling using an LDA was performed on the miss-classified tweets of the RoBERTa model to gain insight on how to further improve model accuracy.
\section*{Nomenclature}
\begin{center}
\begin{tabular}{ll}
 SVM & Support Vector Machine.  \\ 
 COVID-19 & Coronavirus Disease-19.  \\  
 NLP &Natural Language Processing. \\ 
 BERT & Bidirectional Encoder. \\
      & Representations for Transformers.\\
 RoBERTa & Robustly Optimized BERT\\
         &  Pre-training Approach.\\
 UGC & User Generated Content.\\
 LSTM & Long Short Term Memory.\\
 Bi-LSTM & Bidirectional-LSTM.\\
VADER & Valence Aware Dictionary and\\
      & sEntiment Reasoner.\\
 LDA & Latent Dirichlet Allocation.\\
 AI & Artificial Intellegence.\\
 NPI & Non-pharmaceutical interventions.\\
 ABSA & Aspect-based Sentiment Analysis.\\
 NB & Na\" {\i}ve  Bayes.\\
 VAI & Vaccine Acceptance Index.\\
 TF-IDF & Term Frequency Inverse Document \\
        & Frequency.\\
 RF & Random Forest.\\
 NSP & Next Sequence Prediction.\\
 MLM & Masked Language Modelling.\\
 WandB & Weights and Biases.\\
\end{tabular}
\end{center}
\vspace{1cm}
\section{Introduction}
The still ongoing COVID-19 pandemic, which represents the most significant healthcare emergency in recent times, has had a shattering effect all over the globe - both physically and psychologically \cite{bonell2020harnessing}. Many NPIs, such as wearing masks, washing hands regularly, and maintaining social distancing, can help reduce the spread of the virus and have been, indeed, effective in mitigating the infectious outbreak.\cite{bonell2020harnessing}\\\\ However, they are not sustainable in the long term, both in terms of acceptability and psychological and economic impact. Also, they may not fully eradicate the disease. In this context, pharmaceutical interventions, including drugs and vaccination, can play a very crucial role in combating the infection and immunization could potentially eradicate this virus.\\\\ Several government agencies have worked closely with public and private organizations worldwide to provide the scientific community with the necessary resources to work toward the development of vaccines and drugs that would protect against COVID-19 infection, as well as mitigate the severity of symptoms arising from COVID-19 infection in the elderly and people with co-morbidities\cite{Ball2016}. While drug discovery and vaccine development and roll-out are, in general, long and complicated processes, often taking an average of 10 to 15 years for development to be completed and approval to be finalized, the prompt development of the pharmacological compounds and vaccines against COVID-19 has been facilitated by several years of past basic and translational research.\cite{Ball2016}\\\\ A global research effort coupled with novel technological advancements has allowed faster ways to manufacture drugs and vaccines while extensive funding has allowed firms to run multiple trials in parallel, thereby, expediting the process enormously ~\cite{Ball2016}. Specifically concerning immunization, according to WHO, by August 2022, there were 198 COVID-19 vaccine candidates in pre-clinical development and 170 in clinical development \cite{COVID-WHO}, but even once a vaccine has been developed and manufactured, challenges are not ended yet. The implementation of a mass immunization campaign may present, indeed, organizational and logistic hurdles and having to face vaccine hesitancy \cite{COVID-WHO}.\\\\ For instance, in South Africa, the national vaccination program against COVID-19 commenced on 17th February 2021 \cite{COVID-vacc2021}. The roll-out of the vaccine strategy in South Africa was implemented in a three-phase approach: first by vaccinating the most vulnerable population such as front-line healthcare workers and, then, by catering to other essential workers, people in congregated settings, persons over 60 years old, people over 18 years old with co-morbidities, and, finally, the population over the age of 18. The goal was to vaccinate at least $67\%$ of the population by the end of 2021 \cite{ doi:10.1080/14760584.2021.1949291}.\\\\ As of May 30, 2022, around $50.03\%$ of adults in the country had had at least one COVID-19 vaccination. Gauteng leads other provinces in terms of the number of jabs administered (over ten million), followed by KwaZulu-Natal with more than five million vaccinations. It is quite apparent that most populations in different provinces have not yet been fully or partially vaccinated \cite{COVID-WHO}.\\\\ The lack of willingness of the public to get vaccinated against COVID-19 is a matter of great concern to both health scientists and workers in the field of public health. After the initiation of the vaccination roll-out process, the public's opinions and emotions have become quite diverse. Different studies have been conducted all over the world with the aim of trying to detect and understand the reasons behind vaccine hesitancy \cite{litvaxhes1}, which represents a complex, multi-factorial phenomenon\cite{COVID-WHO}.\\\\ From these studies, some of the reasons that were identified were erroneous beliefs such as those that the vaccines were produced too quickly without proper research being undertaken, the vaccines were thought to cause cancer and/or infertility, uncertainty regarding the second dose's availability, increased risk of serious side-effects for people with pre-existing conditions/co-morbidities, and possible allergic reactions\cite{litvaxhes2}. Also instrumental in the rapid rise in vaccine hesitancy were the spread and propagation of conspiracy theories and misinformation - which were due to anti-science, political and religious posts on social media persuading users towards adopting an anti-vaccination attitude.\cite{litvaxhes2}\\\\ This rapid flow of multiple sources and types of misinformation drastically slowed down the acceptance of the COVID-19 vaccines. Several opposing opinions further divided the general population into groups of people and created a near-hostile temperament toward the topic of vaccination.\cite{litvaxhes2}\\\\ A previous study done by B. Mellado et al was  published in a paper with the title: Leveraging Artificial Intelligence and Big Data to Optimize COVID-19 Clinical Public Health and Vaccination Roll-Out Strategies in Africa” has shown that ``Big data and artificial intelligence (AI) machine learning techniques and collaborations can be instrumental in an accurate, timely, locally nuanced analysis of multiple data sources to inform CPH decision-making, vaccination strategies and their staged roll-out” \cite{Bruce2021}.\\\\ Therefore, the government and other agencies should analyze people’s sentiments about vaccination campaigns to maximize and optimize their roll-out, collecting available data from different social networking sites. Examples of UGC \cite{{naab2017studies}} include tweets, Facebook status updates, videos, blogs, forum posts, and consumer-produced product reviews, among others. \cite{smith2012does}\\\\ UGC can be mined in order to identify trends and make predictions on a range of diverse subjects and topics, spanning from product launches and sales to political campaigns and elections, natural disasters, infectious epidemics, and pandemics. Concerning the latter topic, there are several studies where UGC has been used to understand people's opinions about the Coronavirus and its spread, government measures taken to control its spread, and the development and administering of vaccines \cite{puri2020social}.\\\\ However, to the best of our knowledge, the public's hesitation associated with getting vaccinated against COVID-19 has been investigated mainly in the Global North, but, to a lesser extent, in the Global South.This is very apparent if one considers that by the 9th July 2021, the share of people that have been partially or fully vaccinated per continent where all under 50$\%$, with North America and Europe having 44$\%$ and 43$\%$ of its residents receiving at least one vaccination against COVID-19, followed by South America with 34$\%$, which is way ahead of Asia and the Oceania each having respective shares of 25$\%$ and 19$\%$, respectively, and then Africa with a dismal amount of under 5$\%$. \cite{Alam2021}\\\\ The fact that Africa is way behind in terms of vaccination rates as compared to the rest of the world, this further justifies and motivates the importance of this study \cite{Alam2021}. Moreover,  the platform most frequently used for delivering thoughts on the COVID-19 situation, since its emergence until now and especially during the year 2021, was Twitter - hence justifying using Twitter data as opposed to data from other social media platforms for this study. \cite{Alam2021}\\\\ A total of 20 related works were analyzed, with the 6 most relevant papers mentioned in the upcoming literature review section, with each using either NLP techniques and/or ML methods in order to probe the public's sentiments towards certain pandemic-related topics such as vaccination and lockdown measures through user comments on one or more social media platforms extracted within or from one country/continent or amongst several countries/continents, with the intention of guiding policy-makers in making decisions, given the devastating effect of the pandemic.\\\\ Most studies exclusively used automated labelling methods in their sentiment analysis, while some included both manual and automated labelling in their experiments. The machine learning models that were commonly used included state-of-the-art models such as BERT, classical models such as SVM and novel recurrent neural networks such as LSTM/Bi-LSTM.\\\\ All these related works showcased the power of sentiment analysis and the potential prowess of using NLP techniques in conjunction with machine learning methods in the research in extracting meaningful conclusions pertaining to people's feelings/ opinions towards a particular topic - which can be used in future studies to create more sophisticated models that would help policymakers in making decisions during a pandemic or public health crises. No studies exclusively used manual labelling in their research, while some used partial manual labelling and others using ABSA with relatively good results.\\\\ However, there are many limitations to these studies involving data bias given the intrinsic characteristics of social media users being young and from more urbanized areas, as well as model bias given the choice of keyword selection. Moreover, other limitations arise from the tremendous amount of time manual labelling takes as opposed to automated labelling, class imbalance, dataset size and characteristics, as well as conflict from subtle deviations in terms of agreement with the choice of definition for vaccine hesitancy and the accompanying sentiment labels, along with the method of pre-processing and the rules used in the labelling process.\\\\ With these observations in mind, this study explored vaccine hesitancy in South Africa, which is in the Southern Hemisphere, using Twitter data as a source of public opinion. More specifically, the aim was to quantify and qualify the public’s willingness to be vaccinated in order to develop an AI model that would be able to detect the presence of vaccine hesitancy and track its dynamics, thereby, paving the path to an AI-mediated response to a global health crisis. This would allow for a faster, more efficient, implementation and deployment of disaster management systems for the detection, mitigation, and eradication of infectious pandemics.   
\section{Related Work}
In 2020,  M.B. Mutanga and A. Abayomi used Twitter data from South Africa and identified issues relating to the pandemic using an LDA, which they showcased in a paper entitled: ``Tweeting on COVID-19 pandemic in South Africa: LDA-based topic modelling approach." From the LDA analysis, some topics that were being discussed were identified pertaining to the sale and consumption of alcohol, lockdown, daily rates of infection, police brutality, 5G radiation causing COVID-19 and vaccines, as well as conspiracy theories. These topics were an illustration of the attitudes and perceptions the citizens had towards the topic of vaccines. The findings also revealed people’s resistance to measures that affect their economic activities, and their unwillingness to take tests or vaccines as a result of fake news and conspiracy theories \cite{Mutanga2020}.\\\\ The study was very comprehensive but is limited given that as the COVID-19 pandemic continues its offence and new sources of damage and opportunities are being found, future work needed to be inclusive of extracting the emotion behind the sentiments from the collected tweets - in order to investigate the evolution of the public's opinions with time before and after certain remarkable events. Testing of additional topic extraction algorithms, including a combination of NLP techniques and machine learning methods toward an automatic classification and prediction of diverse factors relating to the COVID-19 pandemic were not performed in this study \cite{Mutanga2020}.\\\\In 2022, a paper entitled: ``Sentiment analysis tracking of COVID‑19 vaccine through tweets," by A. Sarirete et al., people's sentiments towards vaccination during the pandemic from tweets that were scraped via the use of the TAGS tool from Twitter users from all over the world were investigated, using a hybrid approach, which combined the use of linear, probability and/or decision tree classifiers with a statistical-, semantics-  and/or a dictionary-based approach. In other words, the hybrid approach uses NLP techniques in conjunction with ML methods, and in this case, were applied in order to classify text and extract the degree of vaccination hesitancy towards COVID-19 vaccines in general\cite{Saritete2022}\\\\ From the corpus analysis, emojis and words related to a sentiment were identified. The frequency of these keywords was recorded and each tweet was classified based on the keyword frequency using the aforementioned machine learning models. It was found that the tweets could be separated into positive and negative sentiments, with a dominance towards the negatives. Although,  several tweets were collected, analyzed and classified based on keywords frequency, manual labelling was absent and more testing is needed on tweets using machine learning techniques to compare the results with the NLP techniques, and generalizing the algorithm to different hashtags and other applications \cite{Saritete2022}.\\\\ In 2021, a paper entitled: ``Sentiment Analysis of COVID-19 Vaccine Perception Using NLP," by M.A.  Mudassir, Y. Mor, R. Munot et al., the sentiments of the people residing in India with regards to the COVID-19 vaccine were analyzed. The paper used three different classification models i.e., TextBlob, VADER, and ABSA to perform the sentiment analysis on English tweets that were posted by users in India and then chose the best deep learning model after comparing their results based on F1-score and test accuracy\cite{Mor2021}.\\\\ TextBlob and VADER are commonly used automated labelling algorithms, while ABSA is an ML that finds and attributes sentiment to aspects, features, and topics that it has categorized within the body of text - more in line with a human perspective used when manually labelling text.\\\\ In this study, 2000 or 10 $\%$ of the tweets in the dataset were manually labelled and tested on the three different models. The model with the highest accuracy was chosen and rest of the tweets were labelled using this model. It was found that ABSA produced the best result out of the other models due to its ability to focus on the specified aspects enhanced by the Attention based Transformer model and it was argued that ABSA should be used more frequently in sentiment analysis studies tasks which have a narrow focus rather than general purpose models\cite{Mor2021}.\\\\ The results of this study showed that the insights gained from the ABSA model were more detailed and descriptive than other techniques that fail to give a more than a general overview of sentiment - however it is a notably a significantly slower method, which will need to be investigated in future studies. Thus, this study illustrates the advantages of using other methods of text classification used in the training phase such as manual labelling in conjunction with ABSA, instead of solely relying on automated labelling methods \cite{Mor2021}.\\\\ In 2021, a paper entitled: ``Dynamic assessment of the COVID-19 vaccine acceptance leveraging social media data," by L. Li and J. Zhou et al, over 29,000,000 vaccine-related tweets from 08 August to the 19th April 2021 were collected and quantified using a VAI, which they computed based on opinion classifications identified with the help of NLP techniques and provided them with a quantitative metric to show the level of vaccine acceptance across different geographic scales in the U.S. Text classification was either automated and performed using TextBlob and VADER or manually labelled into one of three classes i.e. positive, negative or unrelated \cite{Li2021}.\\\\ A fixed sample of 20000 unique tweets from the collected tweets that were most frequently re-tweeted were manually labelled according to specific labelling criteria, which were based on the CDC strategy in order to consolidate confidence in COVID-19 vaccines, whereby 10$\%$ of this dataset was chosen for the testing sample. A total of 9 candidate models were selected and then trained and tested on the aforementioned tweets and the best model in terms of F1 score and accuracy was selected after an extensive grid search was performed in order to obtain the model-specific set of hyperparameters whose values have been optimized to provide the best possible performance \cite{Li2021}.\\\\ The TF-IDF + RF that was trained on an augmented training set obtained the best overall performance and hence was applied to the entire dataset in subsequent steps. A classification was assigned to each tweet and used to compute a user-based vaccine acceptance measure. Different VAI measures were constructed for national-level, state-level and country-level analysis, respectively \cite{Li2021}.\\\\ At the national level, it showed that the VAI transitioned from negative to positive in 2020 and stayed steady after January 2021 - which was supported by national vaccination rates over that time interval - and re-iterated via a comprehensive analysis of the state- and county-level data. The paper discussed information characteristics that enabled a consistent method of estimation of the VAI \cite{Li2021}.\\\\ The findings supported the use of social media to understand opinions and to offer a fast and inexpensive way to assess vaccine acceptance.– which is also relevant. Therefore, future work could consider using NLP and machine learning tools trained in other languages and integrating data from surveys or models to complement the social media estimation, as well as considering the generalizability of this research framework by applying it to investigate the vaccine acceptance on other types of vaccine and in a broader geographical scale, such as the vaccine acceptance over HPV vaccine and flu vaccine in different countries \cite{Li2021}.\\\\In 2021, a paper entitled: ``Applying Machine Learning to Identify Anti-Vaccination Tweets during the COVID-19 Pandemic", by Quyen G. and Kien G. et al, the performance of various different NLP models i.e., BERT, NB, SVM and Bi-LSTM networks with pre-trained GLoVe embeddings in identifying anti-vaccination tweets published during the COVID-19 pandemic were evaluated \cite{Quyen2021}.\\\\ From the 1st of Jan up until the 23rd of August 2020, ~150,000,000 tweets from all over the world were collected using a Twitter Stream API which allowed public access to a one percent sample of the daily stream of Twitter. After removing all non-English tweets and re-tweets, $\approx$75,000,000 tweets were left behind and used for training and testing \cite{Quyen2021}.\\\\ A systematic random sampling method was used to select 20,854 tweets from 1,474,276 tweets for automated labelling. This sampling method made sure that tweets from across different time intervals during the pandemic were chosen. Tweets were labelled as either “anti-vaccination” or “other”  as the model was aimed to use for stance analysis using stance analysis, in which a tweet is determined to be in favour or against a target \cite{Quyen2021}.\\\\ The optimal model performance on the test set for the BERT model was: accuracy = 91.6$\%$, precision = 93.4$\%$, recall = 97.6$\%$, F1 score = 95.5$\%$, and AUC = 84.7$\%$. From this result along with the other optimized model performances, it was concluded that the BERT models had outperformed all of the other models across all metrics and that given its excellent performance is viable as an identifier of anti-vaccination attitudes from tweets \cite{Quyen2021}.\\\\ However, since stance analysis was used, which is different from sentiment analysis in which a tweet is classified as positive or negative, a negative tweet may not mean anti-vaccine while a positive tweet may not mean pro-vaccine and moreover, only two classes were chosen, which both may have served to inflate the model to such high-performance values. Moreover, it may be possible that BERT has a high correlation with tweets labelled with Textblob and Vader, which needs to be investigated \cite{Quyen2021}.\\\\ Hence, this study should be repeated and cross-checked with the results of similar studies, as well as to check if such a correlation exists and also to compare results against a manually labelled dataset, as well as performing sentiment analysis on the dataset using the same labels and then extending the number of classes to three, in order to verify whether or not this model is reliable as a tool for identifying anti-vaccination attitudes across the globe towards the COVID-19 vaccines \cite{Quyen2021}.\\\\ In 2021, a paper entitled: ``Deep Learning-Based Sentiment Analysis of COVID-19 Vaccination Responses from Twitter Data", by K.N. Alam and Md.S. Khan et al, the authors used a Kaggle dataset called "All COVID-19 Vaccines Tweets" consisting of 125906 vaccine-related tweets from across the globe to train LSTM and Bi-LSTM sentiment classifiers, whereby the tweets were labelled by VADER and not by hand \cite{Alam2021}.\\\\ However, portions of the labelled tweets were assessed and if the label didn't match the sentiment that a human would have given it based on some rules, the cut-offs used for polarity identification were adjusted until more tweets had automated labels that agreed with their ascribed manual labels. This process was iterative and was done until the optimal or near-optimal cut-offs for the three sentiments are found\cite{Alam2021}.\\\\ From the datasets, 125,906 tweets were analyzed using the lexicon-based VADER and separated into three parameters: positive, negative, and neutral. It was found that neutral tweets formed the majority; the negative reactions were lowest in frequency, indicating that fear and unrest related to COVID-19 vaccination procedures were still at large\cite{Alam2021}.\\\\ LSTM and Bi-LSTM models were trained and tested on this dataset, in which The LSTM architecture showed 90.59$\%$ accuracy, and the Bi-LSTM model showed 90.83$\%$ accuracy, and both models showed good prediction scores in the precision, recall, F-1 scores, and confusion matrix calculation\cite{Alam2021}.\\\\ Upon other analyses, it was found that people's reactions towards vaccines, the words “first dose”, “second dose”, “Moderna”, “Pfizer”, “Bharat BioNTech”, “death”, “emergency”, “Covishield” and “clinical trial” were very commonly used by twitter users in Canada and India, along with alarming words like “blood clot”, “feel” and “trial”\cite{Alam2021}.\\\\ Furthermore, from January 21 to the end of February 21, the number of tweets related to vaccines was fewer than 500; from March 21 it rose to nearly 3000, indicating that people were very excited about the vaccines after the completion of the clinical trials and the vaccines were to be administered in large numbers. Then, from March 21 to the present, tweets regarding COVID-19 vaccines had fluctuated from 1000 to 2500 per month, which indicated people’s emotions about them had greatly transformed \cite{Alam2021}.\\\\ This is an example of another study that showed the power of using NLP techniques alongside machine-learning methods in probing people’s vaccination attitudes, as well as their underlying characteristics, across the globe towards the COVID-19 vaccines \cite{Alam2021}.
\section{Experimental Procedure}
A total of 30000 tweets were collected using the Twitter Research License. The extraction focused on hashtags related to vaccines and vaccination over a time period spanning from the 5th March 2020 - when COVID-19 was first identified in South Africa - to the 24th November 2021 when the Omicron variant was first detected in South Africa. Duplicate tweets were removed, leaving 27069 unique tweets. In this study, two distinct pre-processing methods were used: corpus-based and semantics-based methods - each having their own unique emoji dictionaries.\\\\ In the corpus-based or lexical pre-processing method, contractions were removed and replaced by their full forms, uppercase text was lower-cased, integers were removed, hashtags were removed, hyperlinks were replaced by the word `url', @mentions were replaced by the term `atUser', repetitions of emojis, as well as all punctuation marks were removed. Thereafter, using a pre-defined emoji lexicon, relevant emojis were replaced by their physical descriptions in words, while other emojis not thought to convey any sentiment were discarded. This was followed by the replacement of common slang terms with their formal definitions, using a slang-term lexicon. The last step was, then, tokenisation using the TweetTokenizer from the NLTK database.\\\\ In the semantics-based pre-processing method, the same afore-mentioned procedure was followed with a few differences i.e. all punctuation marks and integers were not removed, upper-cased text was not lower-cased, each @mention was replaced by the word `Name' followed by an integer denoting its position relative to other @mentions in each tweet. Furthermore, the emoji lexicon was revised in order to describe the context of the emotion inherent in the emojis and the dictionary of slang terms was extended to include slang terms meaning vaccine or vaccinated such as `vaxxed'  and `vaxx'. Certain hallmark pre-processing steps were not followed i.e., the removal of stop-words, lemmatization, and/or stemming. This was deliberately done in order to preserve the context of the tweets and, thus, the core sentiment. See Tables 4 and 5  under Section II of the Appendix for more details.\\\\Topic Modelling was then performed. The procedure was as follows: The dataset was converted to a list of tweets. Thereafter, standard pre-processing was performed, in which hashtags, urls, emojis and punctuation were removed. Bi-gram and tri-gram models were built with functions defined for stopword removal and text lemmatization. Then contractions were replaced with their full form and stopwords removed. Lemmatization was performed, in which the nouns, verbs, adverbs and adjectives were kept. A dictionary was created using ‘id2word’, in which each word was given its own integer I.D. The corpus from the lemmatized data was created in the form of a list of term and term frequency entries. An LDA was then built using the Gensim module lda-model function, whereby the optimal value for the number of topics was determined to be 5 at a coherence value of 0.3707. The top 30 most salient terms in each topic were extracted, topics visualized using the pyLDAvis tool and thereafter, the topics where identified.
\subsection{\textit{Hand-Labelling of Tweets}}
Before we can motivate why we used manual labelling over automated labelling that uses sentiment analysis algorithms, it is useful to consider why automated sentiment analysis is so popular and the challenges that arise in machine learning when using a classification algorithm or when building a specific type of machine learning classification model. Firstly, in both manual and automatic labelling, there are unavoidable factors that will impact the reliability of the labelled data to some extent, thus making any analytic results not directly applicable to real-world problems. The most concerning factors are mentioned below:
\begin{itemize}
    \item Subjectivity of Text
    \item Context and Tone of the statement
    \item Presence of Sarcasm and Irony 
    \item Presence of Negations  
    \item Use of Emojis and Special Characters 
    \item Use of idiomatic expressions  
    \item Use of Colloquialisms and Slang \cite{Medel2017}. 
\end{itemize}
These are all relevant because many misclassifications by an algorithm or classification model arise directly from these factors. Moreover, even though many classification algorithms have been formulated such as TextBlob and VADER \cite{Medel2017}, AI-based algorithms continue to struggle with - or are completely incapable of -  detecting and understanding human emotion, and since tweets contain a strong emotional component, this may give rise to misinterpretation of text and incorrect labelling - when analysed by humans \cite{Medel2018}.\\\\Despite this, the benefit of using automated labelling over manual labelling is 2-fold:
\begin{itemize}
    \item Removal of Text Subjectivity 
    \item Reliable and Realistic Labels \cite{Medel2017}. 
\end{itemize}
Firstly, text subjectivity which arises from the fact that the meaning behind a statement is understood through our own life experiences and unconscious biases that, we, as humans have developed over the years, is no longer an influencing factor. This is obvious, since the tone and context of a piece of text is not considered by an algorithm i.e., it is always objective in its decision making - unlike humans who often will encounter texts that are difficult to classify.\\\\ Secondly, for human beings, labelling text is a long, tiring, and time-consuming process, which is not the case of machines. For example, sentiment analysis algorithms can analyze hundreds of Mbs of text within minutes - while the average human would struggle to label more than 45 tweets in an hour \cite{Medel2017}.\\\\ However, we would like to draw meaningful conclusions from our analysis. So, it is important that the dataset that we used for training and testing NLP classifiers have reliable and realistic labels that are applicable in the real world \cite{Val2020}. Thus, hand-labelling of our dataset is justified in this regard.\\\\Furthermore, even though the precision of these classification algorithms is quite high owing to a consistent sentiment analysis not impacted by subjectivity, the accuracy of the labels from a human perspective would be incredibly low \cite{Petro2021}. Even though human beings would occasionally disagree on the correct label of a text in a large enough dataset, they are still much better apt at understanding the meaning behind the text \cite{Petro2021}.\\\\ It is possible to mitigate the effect of subjectivity when hand-labelling text. This is especially useful and the findings of sentiment analysis would be relevant and important to policymakers. However, one can extend this mitigation by creating a fixed and unchanging bias that is used during the manual labelling process. This is not easy but the more defined the subject matter of the text that we are analysing, the more consistent and dependable the dataset will be, once labelling is complete \cite{Medel2017}. This is imperative and aligns with the aim of this study, which is to create a machine learning model that would be able to accurately predict sentiments pertaining to vaccine hesitancy in order to guide current policy-makers during a pandemic.\\\\ The hand-labelling of the dataset was done by several persons in the team using a strict, clear, and consistent set of rules to minimise the frequency of disagreements on the correct label for a particular tweet. Such workforce also serves to minimize labelling errors, maximize quality control by checking that the labelling rules used were correctly and consistently implemented amongst the labellers and by finding consensus on difficult-to-label tweets.\\\\ A collection of 30000 tweets were selected to be hand-labelled. A label is ascribed to the tweet based on the opinion of the author towards a particular theme or topic - in this case, vaccination. Each tweet was hand-labelled into one of three sentiment classes i.e., positive, negative, or neutral. The criteria for hand-labelling involved answering a simple question: ``Does the author of this comment approve or disapprove in taking a vaccination shot against COVID-19 and to what extent does he/she agree or disagree?"\\\\ To answer this question, a careful look at the punctuation, grammar, choice of words and symbols as well as tone inherent in the tweets were examined. To make things easier to categorize, easy-to-label tweets were labelled first and difficult-to-label, in which both negative and positive sentiments could be found in the tweet, were left towards the end.\\\\We adopted IBM’s definitions of vaccine hesitancy, which was based on WHO's definition of vaccine hesitancy in this paper, in which a negative sentiment was defined as a refusal to get vaccinated - which is referred to as overt hesitancy, a positive sentiment was defined as a decisive decision to get vaccinated, while a neutral sentiment was defined as a delay towards getting vaccinated i.e. an indecisive temperament towards vaccination - this is referred to as subtle hesitancy \cite{WHO2015}.\\\\ Additionally, a statement in which the author's viewpoint is unclear or unrelated to vaccination is by default labelled as neutral. Hence, we argue that the practice of hand-labelling is superior to automated classification algorithms - which frequently mislabel text that contain certain tones and contexts especially when negations, colloquial slang, emojis, and sarcasm are present, as previously said.\\\\Refer to Table 3 under Section I of the Appendix for examples whereby the afore-mentioned statement is true\\\\ In the next section, we introduce the type of machine learning algorithms/models that we used, briefly discussing the architecture of the models, as well the data processing, training, and testing steps that were involved.
\subsection{\textit{Support Vector Machines}}
The Support Vector Machine (SVM) algorithm is a popular algorithm amongst supervised machine learning algorithms, which can be utilized for classification purposes as well as for solving regression problems. In our model, feature extraction and the ``Bag of Words" model were implemented in the post pre-processing steps with vectorisation being performed by the TFidfVectorizer on the text samples, while the labels were label$-$encoded using the Label$-$Encoder function from sklearn. The max number of features, in this case, was chosen to be 5000. This procedure is the preferred data preparation technique for SVMs, within the context of NLP\cite{G4G2020}.
\subsection{\textit{LSTM and Bi-LSTM}}
Both LSTMs and Bi-LSTMs are recurrent neural networks (RNNs). LSTM-based models make use of both Long-Term Memory (LTM) and Short-Term Memory (STM) in simplifying calculations through the application of gates i.e., Forget Gate, Learning Gate, Recall Gate and an Output Gate \cite{Kumar2020}. Owing to its bi-directionality, the Bi-LSTM is, in general, considered to be more effective in the deep learning process as compared to LSTMs \cite{Kumar2020}.\\\\The architecture of both models was chosen to be the same, for this study, i.e., both the LSTM and Bi-LSTM models consisted of an Input layer followed by an Embedding layer, Dense layer, two LSTM or bi-LSTM layers, another Dense layer and finally an Output Dense layer, whereby each individual layer is separated by a Dropout layer.\\\\The activation function for all the Dense layers was chosen to be `relu', except for the Output layer with activation function, `softmax'. The model was compiled with a loss function of `categorical crossentropy'. The argument for stacking LSTM or Bi-LSTM layers on top of each other is to allow for a greater model complexity \cite{Kumar2020}.\\\\The labels for the targets, were not label-encoded as was performed in the case of SVM into categorical variables of varying weight-age but made into categorical variables that each carry equal weighting. Our choice of embedding technique was feature extraction with a maximum number of features set to 2000.
\subsection{\textit{BERT and RoBERTa}}
Over the past years, supervised models have shown consistently better results than unsupervised models, until the introduction of a new pre-trained BERT text attachment model, which enabled unprecedented precision of results in many automated word processing tasks. This model replaced the widely known ``word2vec" model in prevalence, becoming the industry standard. This is the motivation for the use of BERT in the study. BERT-base-cased was chosen, since it does not lowercase the sample text, thus preserving tone and context, and will take less computational power and time to train than BERT-large-cased. Soon after the construction of BERT, Robustly optimized BERT approach, RoBERTa was formed. Since RoBERTa is a retraining of BERT with improved training methodology, more data and computational power, in which the training procedure is improved whereby RoBERTa removes the Next Sentence Prediction (NSP) task from BERT’s pre-training and, instead, it introduces dynamic masking, this model i.e. RoBERTa-base was also chosen in the study. Both the BERT-base-cased and RoBERTa-base models were chosen for fine-tuning and both trained and evaluated on our dataset, with results then being compared to pre-selected pre-trained models evaluated on our dataset.
\subsection{\textit{Model Hyper-parameters Used}}
In this study, all the machine learning models underwent extensive hyperparameter tuning using Bayesian optimization. The hyperparameters chosen for tuning in the case of the SVM were the cost function, C, $\gamma$, and the kernel. The hyperparameters chosen for tuning for both the LSTM and bi-LSTM models were the dropout rate, learning rate, weight decay, batch size, dense units, embedded dimensions, hidden dimensions, number of epochs and choice of optimizer.\\\\ The hyperparameters chosen for tuning for both BERT-base-cased and RoBERTa-base were the learning rate, batch size and number of epochs - but not the weight decay, which was set to zero. Given the uniqueness and complexity of the data-set and subject matter, the slight shift away from a balanced dataset, the small size of the dataset, as well as the non-typical method of labelling that was used, the overall and individual F1-scores were chosen as the defining measures for which the models could be assessed. For the Model-specific hyperparameters and their pre-selected ranges, please see Tables 6 and 7 of the Appendix under Section-III. Next, we will discuss model performance.\\
\section{Results and Discussion}
\subsection{\textit{Machine Learning Models}}
\begin{table}[!ht] 
\centering
\begin{tabular}{|p{0.06\textwidth}|p{0.0225\textwidth}|p{0.0225\textwidth}|p{0.0225\textwidth}|p{0.0225\textwidth}|p{0.0225\textwidth}|p{0.0225\textwidth}|p{0.0225\textwidth}|p{0.0225\textwidth}|}
		\hline
		\multicolumn{9}{|c|}{\tiny{Model Classification Report}}\\
		\hline
		\multicolumn{9}{|c|}{\tiny{SVM}}\\
		\hline
		&  \multicolumn{4}{|c|}{\tiny{ Semantics Approach }} &  \multicolumn{4}{|c|}{\tiny{ Lexical Approach }}   \\
		\hline
		\tiny Class & \tiny{Neg} & \tiny{Neu} & \tiny{Pos} & \tiny{All} &  \tiny{Neg} & \tiny{Neu} & \tiny{Pos} & \tiny{All} \\
		\hline
		\tiny Precision & \tiny 53 & \tiny 49 & \tiny  62 &  \tiny 54 & \tiny 53 & \tiny 49 & \tiny 62 &  \tiny 54 \\
		\hline
		\tiny Recall & \tiny 56 & \tiny 51 & \tiny 56  &  \tiny 54 & \tiny 56 & \tiny 51  & \tiny 56  &  \tiny 54\\
		\hline
		\tiny F1-score & \tiny 54  & \tiny 50  & \tiny 59 &  \tiny 54 & \tiny 54 & \tiny 50 & \tiny 59 &  \tiny 54\\
		\hline
		\tiny Accuracy & \multicolumn{3}{c|}{\tiny{}} & \tiny{54} & \multicolumn{3}{|c|}{\tiny{}} & \tiny{54}\\
		\hline
		\multicolumn{9}{|c|}{\tiny{LSTM}}\\
		\hline
		&  \multicolumn{4}{|c|}{\tiny{ Semantics Approach }} &  \multicolumn{4}{|c|}{\tiny{ Lexical Approach }}   \\
		\hline
		\tiny Class & \tiny{Neg} & \tiny{Neu} & \tiny{Pos} & \tiny{All} & \tiny{Neg} & \tiny{Neu} & \tiny{Pos} & \tiny{All} \\
		\hline
		\tiny Precision & \tiny 49 & \tiny 46  & \tiny 50 &  \tiny 48 & \tiny 47 & \tiny 46 & \tiny 58 &  \tiny 50\\
		\hline
		\tiny Recall & \tiny 43  & \tiny 41  & \tiny 63 &  \tiny 49 & \tiny 58 & \tiny 42 & \tiny 44 &  \tiny 48\\
		\hline
		\tiny F1-score & \tiny 46 & \tiny 43 & \tiny 56 &  \tiny 48 & \tiny 52 & \tiny 44 & \tiny 50 &  \tiny 49 \\
		\hline
		\tiny Accuracy & \multicolumn{3}{c|}{\tiny{}} & \tiny{49} & \multicolumn{3}{|c|}{\tiny{}} & \tiny{48}\\
		\hline
		\multicolumn{9}{|c|}{\tiny{Bi-LSTM}}\\
		\hline
		&  \multicolumn{4}{|c|}{\tiny{ Semantics Approach }} &  \multicolumn{4}{|c|}{\tiny{ Lexical Approach }}   \\
			\hline
		\tiny Class & \tiny{Neg} & \tiny{Neu} & \tiny{Pos} & \tiny{All} & \tiny{Neg} & \tiny{Neu} & \tiny{Pos} & \tiny{All} \\
		\hline
		\tiny Precision & \tiny 50  & \tiny 44 & \tiny 52  &  \tiny 49 & \tiny 50 & \tiny 49 & \tiny 59 &  \tiny 53 \\
		\hline
		\tiny Recall & \tiny 46  & \tiny 50 & \tiny 57  &  \tiny 51 & \tiny 49 & \tiny 52 & \tiny 57 &  \tiny 52\\
		\hline
		\tiny F1-score & \tiny 43  & \tiny 47 & \tiny 54 &  \tiny 50 & \tiny 49 & \tiny 50 & \tiny 58 &  \tiny 52\\
		\hline
		\tiny Accuracy & \multicolumn{3}{c|}{\tiny{}} & \tiny{51} & \multicolumn{3}{|c|}{\tiny{}} & \tiny{52}\\
		\hline
		\multicolumn{9}{|c|}{\tiny{BERT}}\\
		\hline
		&  \multicolumn{4}{|c|}{\tiny{ Pre-trained }} &  \multicolumn{4}{|c|}{\tiny{ Fine-tuned }}   \\	
            \hline
		\tiny Class & \tiny{Neg} & \tiny{Neu} & \tiny{Pos} & \tiny{All} & \tiny{Neg} & \tiny{Neu} & \tiny{Pos} & \tiny{All} \\
		\hline
		\tiny Precision & \tiny 48 & \tiny 43  & \tiny 46 &  \tiny 46 & \tiny 60 &  \tiny 57  & \tiny 64 & \tiny 60\\
		\hline
		\tiny Recall & \tiny 43  & \tiny 46   & \tiny 49  &  \tiny  46  & \tiny 64 & \tiny 51  & \tiny 69 & \tiny 61 \\
		\hline
		\tiny F1-score & \tiny 46 & \tiny 45   & \tiny 48   &  \tiny 46  & \tiny 62 & \tiny 54 & \tiny 66 & \tiny 60 \\
		\hline
		\tiny Accuracy & \multicolumn{3}{c|}{\tiny{}} & \tiny{46} & \multicolumn{3}{|c|}{\tiny{}} & \tiny{61}\\
		\hline
		\multicolumn{9}{|c|}{\tiny{RoBERTa}}\\
		\hline
		&  \multicolumn{4}{|c|}{\tiny{ Pre-trained }} &  \multicolumn{4}{|c|}{\tiny{ Fine-tuned }}   \\
	    \hline
		\tiny Class & \tiny{Neg} & \tiny{Neu} & \tiny{Pos} & \tiny{All} & \tiny{Neg} & \tiny{Neu} & \tiny{Pos} & \tiny{All} \\
		\hline
		\tiny Precision & \tiny 51  & \tiny 42  & \tiny 45 &  \tiny 46  & \tiny 61  & \tiny 58  & \tiny 66 &  \tiny 62 \\
		\hline
		\tiny Recall & \tiny 47  & \tiny 49  & \tiny 52   &  \tiny 48  & \tiny 65 & \tiny 52  & \tiny 58  &  \tiny 61 \\
		\hline
		\tiny F1-score & \tiny 49   & \tiny 45   & \tiny 48 &  \tiny 47  & \tiny 63 & \tiny 55  & \tiny 67 &  \tiny 61 \\
		\hline
		\tiny Accuracy & \multicolumn{3}{c|}{\tiny{}} & \tiny{48} & \multicolumn{3}{|c|}{\tiny{}} & \tiny{61}\\
		\hline
	\end{tabular}
\caption{\textbf{\scriptsize{{\underline{A Summary of Optimised Model Performance.}}}}}
\end{table}
Applying hand-labeling of the tweets, the distribution of sentiments in our dataset were as follows: 31.7$\%$ positive; 36.1$\%$ neutral; 32.2$\%$ negative. There is not a dominant sentiment, and each sentiment is distributed equally within the sentiment population.\\Table 1, above, shows a summary and comparison of optimised model performance for the various models. In terms of model performance, the LSTM model achieved an overall precision of 48$\%$ and overall accuracy of 49$\%$, with a combined F1-score of 49$\%$ using the semantics pre-processing, with a similar result being achieved with an overall precision of 50$\%$, an overall accuracy of 48$\%$, and a combined F1-score of 48$\%$, on the lexical pre-processing approach. As expected, the Bi-LSTM models performed better than the LSTM models, with the Bi-LSTM model achieving an overall precision of 49$\%$, an overall accuracy of 51$\%$, and a combined F1-score of 50$\%$ on the semantics pre-processing approach, but a significantly better result on the lexical pre-processing approach with a higher overall precision of 53$\%$, and an overall accuracy of 52$\%$, yielding an F1-score of 52$\%$. Furthermore, the SVM model achieved identical results for both pre-processing methods with an overall precision of 54$\%$, an overall accuracy of 54$\%$, and a combined F1-score of 54$\%$. It is clear that the SVM model produced results that were better than both pairs of LSTM and Bi-LSTM models, which may sound counter-intuitive at first glance, but can be explained as follows: feature embeddings as used in the SVM model generally perform better than word embeddings as were used in the LSTM/bi-LSTM models in the context of NLP.\\\\Results of pre-selected BERT and RoBERTa pre-trained models served as a comparison to the performance of our respective fine-tuned models. Since, their classification measures were much lower than those of our fine-tuned models, we showed that the manual labelling of the vaccination hesitancy data-set was a novel approach and confirmed that the dataset is more complex than simple sentiment analysis on texts in categorising them into positive, negative or neutral labels based on the overall emotion inherent in the text samples and not on a specific topic. The pre-trained models chosen was a RoBERTa twitter sentiment model by the Cardiff NLP Group \cite{BERTpre}. This pre-trained RoBERTa model was trained on $\approx$ 58M tweets and was fine-tuned for sentiment analysis using the TweetEval benchmark using positive, negative and neutral labels. The pre-trained BERT model that was chosen is a fine-tuned version of a multilingual BERT-base multilingual sentiment model by the NLP-Town Group \cite{RoBERTapre}, which is a  model for sentiment analysis using positive, negative and neutral labels, trained to classify product reviews on a scale of 1 to 5 stars, in English, German, French, Spanish, and Italian. The pre-trained BERT model achieved an overall precision of 46$\%$, and an overall accuracy of 46$\%$, yielding an F1-score of 46$\%$, while our fine-tuned BERT-base-cased model achieved a much better result with an overall precision of 60$\%$, and an overall accuracy of 61$\%$, yielding an F1-score of 60$\%$. By pairwise comparison, the RoBERTa models performed better than the BERT models  with the pre-trained RoBERTa model achieving an overall precision of 46$\%$, overall  accuracy of 48$\%$, yielding an F1-score of 48$\%$, while the RoBERTa model achieved a much better result with a higher overall precision of 62$\%$, and an overall accuracy of 61$\%$, yielding a an F1-score of 61$\%$.\\\\Hence, from these results, one can conclude that  based on the overall weighted F1-score, the best models for this classification problem in decreasing order of optimal performance were fine-tuned Roberta-base, fine-tuned BERT-base-cased, SVM, Bi-LSTM, LSTM, pre-trained RoBERTa-base and lastly pre-trained BERT-base-uncased.
\subsection{\textit{Topic Modelling}}
An LDA was performed on the set of tweets that were miss-classified by the model with the highest efficiency – in this case, RoBERTa base. The top 10 Most Frequent Terms per LDA Cluster Grouping are shown, below, in Table 2. Given these keywords, the topics related to vaccine hesitancy were inferred and described i.e., mass vaccination roll out schemes in terms of availability and service delivery, defiance in response to international travel restrictions that target the non-vaccinated or partially-vaccinated population, safety concerns about severe side-effects from the vaccine, as well as concerns of ineffectiveness of vaccines in preventing the spread of the virus. The close enough 5 topics that are inferred from the LDA in which the clusters were visualized are mentioned, below, in Figure 1:\\
\begin{itemize}
	\item Topic 1 : Inefficient mass vaccination 
	
	\item Topic 2 : Selective air travel restrictions 
	
	\item Topic 3 : Severe side-effects 
	
	\item Topic 4 : Inescapable from illness/death 
	
	\item Topic 5 : Ineffective to COVID 
\end{itemize}
\begin{figure}[ht]
	\centering
	\includegraphics[width=0.3\textwidth,height=0.2\textwidth]{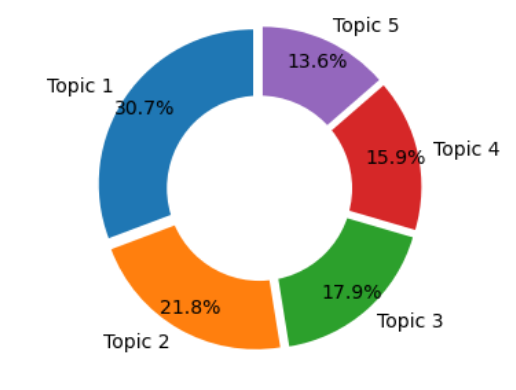}
 \caption{\textit{Distribution of topics in the sample space of tweets miss-classified  by the RoBERTa-base algorithm.}}
\end{figure}
\begin{table}[ht!]
\centering
{\begin{tabular}{ |p{0.01\textwidth}|p{0.215\textwidth}|p{0.215\textwidth}|}
\hline
\footnotesize{} & \multicolumn{2}{|c|}{\footnotesize{Top 10 Most Frequent Terms}}\\
\hline
\footnotesize{T1 } & \multicolumn{2}{|p{0.43\textwidth}|}{\footnotesize{vaccine, government, people, service, shortage, roll-out, manage, slow, help, far}}\\
\hline
\footnotesize{T2 } & \multicolumn{2}{|p{0.43\textwidth}|}{\footnotesize{travel, spread, forced, location, ban, global, spread, rate, cases, news}}\\
\hline
\footnotesize{T3 } & \multicolumn{2}{|p{0.43\textwidth}|}{\footnotesize{pain, afraid, hospital, deadly, risk, serious, allergy, report, approve, poison}}\\
\hline
\footnotesize{T4 }& \multicolumn{2}{|p{0.43\textwidth}|}{\footnotesize{fear, die, spread, infection, real, fast, unsafe, hospitals, symptoms, stuck}}\\
\hline
\footnotesize{T5 }& \multicolumn{2}{|p{0.43\textwidth}|}{\footnotesize{variant, mutation, new, ineffective, dose, booster, second, ongoing, re-infection, partially}}\\
\hline
\footnotesize{All } & \multicolumn{2}{|p{0.43\textwidth}|}{\footnotesize{ vaccination, vaccine, people, covid, health, virus, country, work, today, need}}\\
\hline
\end{tabular}}
\caption{\textbf{\underline{\scriptsize{Top 10 Most Frequent Terms per LDA Cluster.}}}}
\end{table}
\subsection{\textit{Limitation Of Study}}
Overall, the model performance is quite low, with 60-65$\%$, being a very average score and also the range pertaining to the best performance achieved from our set of models. This could be due to a number of reasons i.e., in the case of the LSTM/bi-LSTM models, additional LSTM layers may be needed in order for the model to better grasp the complexity inherent in the dataset; while for the SVM algorithm, it is possible that a different word embedding technique and alternative model other than feature extraction and `Bag of Words' would have resulted in a better performance; while in the case of BERT and RoBERTa, the use of the -LARGE- formulations of these models instead of the -BASE- formulations should be used to improve the performance, and their performances may further be enhanced by incorporating some of the pre-processing steps used in the other non-Transformer models or fine-tuning on other BERT and RoBERTa models trained on similar pandemic-related use-cases for sentiment classification or by using a particular BERT or RoBERTa model with a respective and alternative tokenizer taken from a different BERT and RoBERTa model. 
\section{Conclusion}
In conclusion, the models used were LSTM, bi-LSTM, SVM, BERT-base-cased and RoBERTa-base, whereby their hyperparameters were carefully chosen and tuned using the WandB platform and trained on a hand-labelled dataset containing tweets from South Africa on the topic of vaccine hesitancy. Out of all The machine learning models, excluding the pre-trained and fine-tuned ones, SVM was the best model with an  overall F1-score of 54$\%$, followed by the bi-LSTM with an overall F1-score of 52$\%$ and lastly the LSTM with an overall F1-score of 49$\%$. The best model overall was the fine-tuned RoBERTa-base model with an overall F1-score of 61$\%$, followed closely behind by the fine-tuned BERT-base-case model with an overall F1-score of 60$\%$, where the best model was defined as the model with the highest overall F1-score. From the LDA on the miss-classified tweets of the fine-tuned RoBERTa model, certain types of vaccine hesitancy where identified as topics, which would serve to improve our best model's performance in future studies to better detect vaccine hesitancy and guide policymakers in managing the pandemic. Furthermore, since BERT and RoBERTa are Transformer models, meaning that they can be trained on downstream tasks, further training on additional data-sets with pandemic-related use-cases other than vaccination hesitancy, such as public compliance to other safety measures or the degree of faith in government interventions etc - whose data may originate from other countries around the globe - could essentially pave the way towards a universal tool for early disease detection and the enforcement of public compliance during public health crises or emergencies.
\section*{Appendix}
\section*{Appendix I}
\section*{Hand-Labelling of Tweets}
\begin{table}[ht]
{\begin{tabular}{ |p{0.05\textwidth}|p{0.16\textwidth}|p{0.04\textwidth}|p{0.04\textwidth}|p{0.05\textwidth}|}
\hline
\tiny Cases & \tiny Tweet & \tiny Hand & \tiny VADER & \tiny TextBlob \\
\hline
\tiny & \tiny I guess it's my turn. Let's make a difference. \includegraphics[scale=0.05]{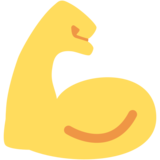} \includegraphics[scale=0.05]{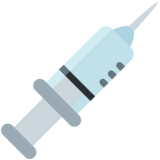} \#VaccineRollOut & \scriptsize + & \scriptsize + & \scriptsize +\\
\cline{2-5}
\tiny Clear-cut & \tiny I refuse to be manipulated in taking an experimental jab. & \scriptsize \textbf{-} & \scriptsize \textbf{-} & \scriptsize 0\\
\cline{2-5}
\tiny & \tiny Should I vaccinate or not? What do you guys think? & \scriptsize 0 & \scriptsize 0 & \scriptsize 0 \\
\hline
\tiny  & \tiny My whole family got jabbed, except for me. All of them had minimal side effects.  & \scriptsize + & \scriptsize 0 & \scriptsize 0 \\
\cline{2-5}
\tiny Border-line & \tiny We must be vaxxed to enter the stadium but not to vote ? \includegraphics[scale=0.03]{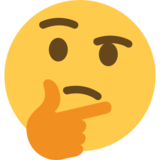} \includegraphics[scale=0.03]{thinking-face.png} \includegraphics[scale=0.03]{thinking-face.png} & \scriptsize \textbf{-} & \scriptsize 0 & \scriptsize 0\\
\cline{2-5}
\tiny  & \tiny Every time I come to vaccinate, the queue, is ridiculously long. \includegraphics[scale=0.03]{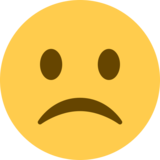} & \scriptsize 0 & \scriptsize \textbf{-} & \scriptsize \textbf{-} \\
\hline
\tiny  & \tiny I have just decided to take the jab. if I die I die. & \scriptsize + & \scriptsize \textbf{-} & \scriptsize 0 \\
\cline{2-5}
\tiny Difficult & \tiny I am fully vaxxed no prob, but a friend took one jab and died. & \scriptsize \textbf{-} & \scriptsize \textbf{-} & \scriptsize 0 \\
\cline{2-5}
\scriptsize  & \tiny Not all people that don't vaxx are anti-vax. Some have health issues. & \scriptsize 0 & \scriptsize 0 & \scriptsize 0 \\
\hline
\tiny & \tiny Let's  Vaccinate to save South Africa !! \includegraphics[scale=0.03]{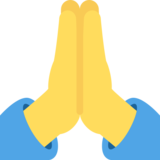} \includegraphics[scale=0.03]{folded-hands.png} & \scriptsize + & \scriptsize + & \scriptsize 0 \\
\cline{2-5}
\tiny Same Text & \tiny Let's Vaccinate  to save South Africa \includegraphics[scale=0.03]{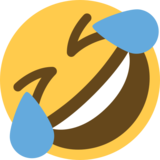} \includegraphics[scale=0.03]{rolling-on-the-floor-laughing.png} \includegraphics[scale=0.03]{rolling-on-the-floor-laughing} \includegraphics[scale=0.03]{rolling-on-the-floor-laughing.png} \includegraphics[scale=0.03]{rolling-on-the-floor-laughing.png} & \scriptsize \textbf{-} & \scriptsize + & \scriptsize 0 \\
\cline{2-5}
\tiny & \tiny Let's  Vaccinate to save South Africa? & \scriptsize 0 & \scriptsize + & \scriptsize 0 \\
\hline
\end{tabular}}
\caption{\textbf{\scriptsize{\underline{Manual versus Automated text labelling.}}}}
\end{table}
Here we highlight the pitfalls of using text classification algorithms over hand-labelling using explicit examples. In Table 3, the four different cases of tweets one would encounter when performing sentiment analysis along with three hand-labelled examples for each case, each corresponding to one of the three sentiment classes i.e., positive (+), negative (-), neutral (0) are provided. The four categories are: clear-cut cases, borderline cases, difficult-to-label tweets and same text tweets. Clear-cut cases correspond to tweets whose sentiment labels are obvious and there is no debate on the validity of its classification - in other words the tweet's polarity is heavily skewed towards a single sentiment type. Borderline cases correspond to tweets that can arguably take on one of two labels i.e., either neutral or positive or alternatively neutral or negative, whereby the author's point of view is debatable. Difficult-to-label tweets are tweets that contain both positive and negative sentiments each with high polarity scores, which makes it difficult to decided on the overall text polarity. Same text tweets are a class of tweets whereby the raw text is identical but differ in the amount of punctuation and/or emojis present in the tweet, which serve to change the message behind the tweet often through the introduction of satire. Two different classification algorithms were selected namely, VADER and TextBlob. These classification algorithms were then given each example tweet and their predicted labels were compared to the manually-classified tweet labels. The results are presented in the table. Overall VADER correctly predicted the labels of 50$\%$ of the tweets, in which 100$\%$ of the clear-cut case examples were classified correctly, while none, or 0$\%$, of the border-line case tweets were classified correctly and only one third, 33$\%$, of the difficult-to-label or the same text tweets were correctly labelled. VADER was able to get 50$\%$ recall for each respective class. Comparatively, TextBlob correctly predicted the labels of a third, or $\approx$ 33$\%$, of all the tweets, in which two thirds, or $\approx$ 67$\%$, of the clear-cut case examples were classified correctly, while none, or 0$\%$, of the border-line case tweets were classified correctly and none, 0$\%$, of the difficult-to-label and the same text tweets were correctly labelled. TextBlob got recalls of 25$\%$ for the positives, 75$\%$ for the neutrals but nothing, 0$\%$, for the negatives. This shows that both classification algorithms perform well on simple clear-cut examples, but become much less efficient in correctly classifying tweets, as the complexity of the tweets increases. Furthermore, given the recall values, it is apparent that VADER is equally good in labelling each sentiment type, while TextBlob strongly favours a neutral label. In both cases, the overall accuracies are very low in comparison to hand-labelling and it is clear that when given same text tweets, the algorithms are unable to identify sarcasm or the nuanced effect of changing punctuation marks i.e., from ! to ?, given that VADER provided a positive label for each sentiment belonging to the same text case, while TextBlob provided all neutral labels. Hence, the table clearly highlights the advantages of manual over automated hand-labelling.
\section*{Appendix II}
\section*{Materials and Methods}
\begin{table}[ht]
{\begin{tabular}{|p{0.015\textwidth}|p{0.185\textwidth}|p{0.195\textwidth}|}
\hline
\multicolumn{3}{|c|}{\scriptsize{Emoji Lexicon}} \\
\hline
\scriptsize{} & \scriptsize{\hspace{0.5cm}Lexical Definition} & \scriptsize{\hspace{0.5cm}Semantics Definition} \\
\hline
\scriptsize \includegraphics[scale=0.8]{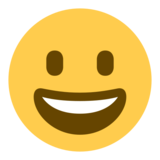} & \scriptsize grinning face & \scriptsize I am happy about this! \\
\hline
\scriptsize  \includegraphics[scale=0.05]{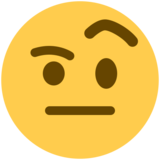} & \scriptsize face with raised eyebrow & \scriptsize I am serious about this. \\
\hline
\scriptsize \includegraphics[scale=0.07]{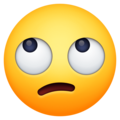} & \scriptsize face with rolling eyes & \scriptsize I do not take this seriously.  \\
\hline
\scriptsize \includegraphics[scale=0.05]{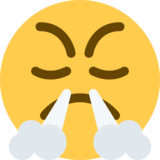} & \scriptsize face with steam from nose & \scriptsize I am angry at this! \\
\hline
\scriptsize \includegraphics[scale=1]{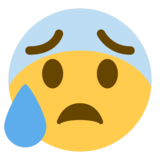} & \scriptsize anxious face with sweat & \scriptsize I do not like this. \\
\hline
\scriptsize  \includegraphics[scale=0.05]{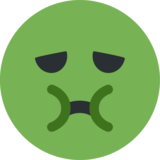} & \scriptsize nauseated face & \scriptsize I am disgusted by this! \\
\hline
\scriptsize  \includegraphics[scale=0.8]{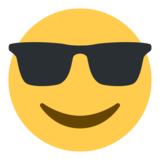} & \scriptsize smiling face with sunglasses & \scriptsize I am proud of this! \\
\hline
\scriptsize \includegraphics[scale=0.05]{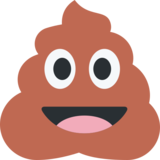} & \scriptsize pile of poo & \scriptsize This is nonsense!  \\
\hline
\scriptsize \includegraphics[scale=0.05]{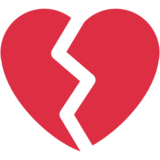} & \scriptsize broken heart & \scriptsize I am sad about this! \\
\hline
\scriptsize \includegraphics[scale=0.05]{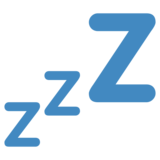} & \scriptsize zzz & \scriptsize I am asleep! \\
\hline
\scriptsize \includegraphics[scale=0.05]{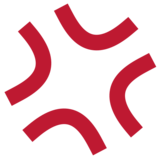} & \scriptsize anger symbol & \scriptsize I am angry about this! \\
\hline
\end{tabular}}
\caption{\scriptsize{\textbf{\underline{A Comparison of the different Emoji Lexicons.}}}}
\end{table}
Here, we show the similarities and differences between the corpus-based and semantics-based approaches. From Table 4, which provides and contrasts the emoji-to-text translation of each approach, it is clear that the two emoji lexicons are very different from each other i.e., while the lexical definition of an emoji is the statement of the physical features of the emoji in words with any punctuation marks, the semantics definition is the statement of the message and emotional intensity attached to that message that is communicated through the use of a particular emoji. The emotional intensity is given by the choice of punctuation marks; in this case only exclamation marks or full stops were used. Not shown in the table are examples of emojis that both methods do not have a definition for and are instead replaced by white-spaces. These emojis were deemed to not carry a sentiment, with an example being the soccer ball emoji.
\begin{table}[ht]
	{\begin{tabular}{ |p{0.085\textwidth}|p{0.093\textwidth}| p{0.091\textwidth}| p{0.098\textwidth}|}
			\hline
			\multicolumn{4}{|c|}{\scriptsize{Pre-processing Rules}} \\
			\hline
			\scriptsize Entity present  & \scriptsize Sample Text & \scriptsize Lexical\hspace{0.5cm} Approach & \scriptsize Semantics Approach\\
			\hline
			\scriptsize URLs & \scriptsize https://host /location & \scriptsize url & \scriptsize `' \\
			\hline
			\scriptsize Slang Terms & \scriptsize vaxxed 2day & \scriptsize vaxxed today & \scriptsize vaccinated today \\
			\hline
			\scriptsize hashtags & \scriptsize ${\#}$word & \scriptsize word & \scriptsize word \\
			\hline
			\scriptsize @mentions  & \scriptsize @USER1 @USER2 & \scriptsize atUser & \scriptsize Name1 Name2 \\
			\hline
			\scriptsize Letter Cases & \scriptsize HeLlO & \scriptsize hello & \scriptsize HeLlO\\
			\hline
			\scriptsize numbers & \scriptsize 30$^{th}$ covid19 2021 & \scriptsize 30 covid & \scriptsize 30$^{th}$ covid19 2021\\
			\hline
			\scriptsize contractions & \scriptsize it’s can’t & \scriptsize it is cannot & \scriptsize it is cannot\\
			\hline
			\scriptsize Repetition & \scriptsize \includegraphics[scale=0.03]{zzz.png} \includegraphics[scale=0.03]{zzz.png} \includegraphics[scale=0.03]{zzz.png} the t!! & \scriptsize zzz the t!! & \scriptsize I am sleeping! t!! \\
			\hline
			\scriptsize Misspelling & \scriptsize i r ded & \scriptsize i r ded & \scriptsize i r ded\\
			\hline
		\end{tabular}}
		\caption{\textbf{\scriptsize{\underline{Pre-processing Rules for the two Approaches.}}}}
\end{table}
\\Here we provide and contrast the pre-processing rules used in the two approaches with explicit examples. The idea behind using two different pre-processing methods was to compare the performance of the models once they were fully trained in order to gain insight into how the models `learn'. Unfortunately, since similar results were obtained by both approaches and a full semantics analysis was not performed, no conclusions or insights could be made on this matter. From Tables 5, it is clear that the pre-processing steps involved in each approach are largely the same, with the exception of how emojis are treated. In both methods, contractions are expanded into their full expressions, spelling errors are not corrected, back-to-back repetitions of emojis within a tweet are discarded leaving behind one of them before it is translated, slang terms are replaced by their formal expressions, hashtags in front of words are removed leaving the words behind, urls and $@$mentions are replaced by some more generic expression, double spaces are contracted into single ones. The minor differences come about in the way that the urls, $@$mentions and slang words related to vaccination are treated by each approach. In the lexical approach, there are no corrections for slang terms pertaining to the word vaccine and its different forms according to the part of speech it adopts in a sentence. The major differences between the two approaches are in the way emojis, punctuation marks, uppercase letters and numerical characters are treated. In the semantics approach, punctuation marks and back-to-back repetitions of punctuation marks are not discarded, uppercase letters were not lower-cased and numerical characters were kept. This is not the case in the lexical approach.
\section*{Appendix III}
\section*{Machine Learning Models}
In Table 6, below, we highlight the hyperparameters of each model; we explicitly show the chosen model-specific hyper-parameters and their associated tuning ranges. The SVM model was chosen to have three hyperparameters i.e., the kernel, gamma, and the cost function, while the LSTM and bi-LSTM models were chosen to have eight hyperparameters i.e, the dropout rate, learning rate, batch size, dense units, embedding dimensions, hidden dimensions, number of epochs, weight decay and the choice of optimizer. The BERT and RoBERTa models were chosen to have three hyperparameters, namely, the number of epochs, dropout rate, and the learning rate. The choice of possible kernels for the SVM models were chosen to be rbf or linear kernels, while the choice of optimizers for the RNN models were chosen to be adam, adamax, rmsprop and SGD. Note that optimizer and weight decay parameters were not chosen as hyperparameters for the transformer models, since it was found that the weight decay was least important with regards to the other three hyperparameters when included as a hyperparameter and since adamW is the standard optimizer used in all BERT/RoBERTa models. Also note that the tuning ranges are much larger than usual, especially in the case of the RNN models, so that an extensive hyperparameter search and optimisation could be performed using the WandB platform. The mode of optimisation was chosen to be Bayes' optimisation.\\
\begin{table}[ht]
	{\begin{tabular}{|p{0.08\textwidth}|p{0.09\textwidth}| p{0.11\textwidth}|p{0.105\textwidth}|}
			\hline
			\tiny{Model} & \tiny{SVM} & \tiny{LSTM/Bi-LSTM} & \tiny{BERT/RoBERTa}\\
			\hline
			\tiny kernel & \tiny rbf, linear & \tiny n/a & \tiny n/a  \\
			\hline
			\tiny $\gamma$  & \tiny (10$^{-4}$; 10$^{-3}$) & \tiny n/a & \tiny n/a \\
			\hline
			\tiny C  & \tiny (10$^{-4}$; 10$^{-3}$) & \tiny n/a  & \tiny n/a \\
			\hline
			\tiny dropout  & \tiny n/a & \tiny (0,0.9) & \tiny n/a  \\
			\hline
			\tiny weight decay  & \tiny n/a & \tiny (0,0.9) & \tiny n/a  \\
			\hline
			\tiny learning rate & \tiny n/a & \tiny (0,0.9) & \tiny (10$^{-6}$,10$^{-4}$) \\
			\hline
			\tiny epochs  & \tiny n/a & \tiny (0,1000)  & \tiny (2,12) \\
			\hline
			\tiny batch size  & \tiny n/a & \tiny (1,1250) & \tiny (8,64)  \\
			\hline
			\tiny dense units  & \tiny n/a & \tiny (1,1250) & \tiny n/a   \\
			\hline
			\tiny embed. dim.  & \tiny n/a & \tiny (1,1250) & \tiny n/a   \\
			\hline
			\tiny hidden dim.  & \tiny n/a & \tiny (1,1250) & \tiny n/a  \\
			\hline
            \tiny optimizer  & \tiny n/a & \tiny Adam,Adamax, RMSprop,SGD & \tiny n/a  \\
			\hline 
		\end{tabular}}
  \caption{\textbf{\scriptsize{\underline{Chosen Model-specific Hyper-parameters to tune.}}}}
	\end{table}
 \begin{table}[ht]
{\begin{tabular}{ |p{0.037\textwidth}|p{0.02\textwidth}| p{0.028\textwidth}| p{0.028\textwidth}|p{0.02\textwidth}|p{0.028\textwidth}|p{0.028\textwidth}| p{0.028\textwidth}|p{0.028\textwidth}|}
\hline
 \tiny Type & \multicolumn{3}{|c|}{\tiny{Semantics-based}} & \multicolumn{3}{|c|}{{\tiny{Lexical-based}}} & \multicolumn{2}{|c|}{\tiny{Fine-tuned}}\\
\hline
    \tiny {Model/\hspace{1cm}hparam} & \tiny{SVM} & \tiny{LSTM}  & \tiny{Bi-LSTM} & \tiny{SVM} & \tiny{LSTM}  & \tiny{Bi-LSTM} & \tiny{BERT} & \tiny{Ro-\hspace{1cm}BERT-\hspace{1cm}a}\\
\hline
\tiny kernel & \tiny rbf & \tiny N/a& \tiny N/a& \tiny rbf & \tiny N/a& \tiny N/a& \tiny N/a& \tiny N/a\\
\hline
\tiny $\gamma$  & \tiny 0.1 & \tiny N/a& \tiny N/a& \tiny 0.1 & \tiny N/a& \tiny N/a& \tiny N/a& \tiny N/a\\
\hline
\tiny C & \tiny4 & \tiny N/a & \tiny N/a& \tiny 4 & \tiny N/a& \tiny N/a& \tiny N/a & \tiny N/a\\
\hline
\tiny drop-\hspace{1cm}out  & \tiny N/a & \tiny 5.863\hspace{1cm}\hspace{1cm}x\hspace{1cm}10{$^{-5}$} & \tiny 3.3354\hspace{1cm}x\hspace{1cm}10{$^{-5}$}  & \tiny N/a& \tiny 1.636\hspace{1cm}x\hspace{1cm}10{$^{-5}$} & \tiny 3.267\hspace{1cm}x\hspace{1cm}10{$^{-5}$} & \tiny 0.1 & \tiny 0.1\\
\hline
\tiny learn. rate & \tiny N/a& \tiny 8.495\hspace{1cm}\hspace{1cm}x\hspace{1cm}10{$^{-5}$} & \tiny 1.515\hspace{1cm}x\hspace{1cm}10{$^{-5}$}  & \tiny N/a& \tiny 5.612\hspace{1cm}x\hspace{1cm}10{$^{-5}$} & \tiny 5.735\hspace{1cm}x\hspace{1cm}10{$^{-5}$} & \tiny 8.841\hspace{1cm}x\hspace{1cm}10{$^{-5}$} & \tiny{5.576\hspace{1cm}x\hspace{1cm}10{$^{-5}$}} \\
\hline
\tiny batch size  & \tiny N/a& \tiny 399 & \tiny 623  & \tiny N/a& \tiny 262 & \tiny 447 & \tiny 32 & \tiny 36 \\
\hline
\tiny dense units  & \tiny N/a& \tiny 804 & \tiny 623  & \tiny N/a& \tiny 776 & \tiny 447 & \tiny N/a& \tiny N/a\\
\hline
\tiny embed. dim.  & \tiny N/a& \tiny 393 & \tiny852  & \tiny N/a& \tiny692 & \tiny 831 & \tiny N/a& \tiny N/a\\
\hline
\tiny hidden dim  & \tiny N/a& \tiny 462 & \tiny 194  & \tiny N/a& \tiny 352 & \tiny 82 & \tiny N/a& \tiny N/a\\
\hline
\tiny epochs  & \tiny N/a& \tiny 96  & \tiny 191  & \tiny N/a& \tiny 66 & \tiny 293  & \tiny 4 & \tiny 7 \\
\hline
\tiny weight decay & \tiny N/a& \tiny 8.944\hspace{1cm}\hspace{1cm}x\hspace{1cm}10{$^{-5}$} & \tiny 8.655\hspace{1cm}x\hspace{1cm}10{$^{-5}$} & \tiny N/a&  \tiny 7.516\hspace{1cm}x\hspace{1cm}10{$^{-5}$} & \tiny 7.729\hspace{1cm}x\hspace{1cm}10{$^{-5}$} & \tiny N/a& \tiny N/a\\
\hline
\tiny opt. & \tiny N/a & \tiny adam & \tiny rms-prop  & \tiny N/a & \tiny adam & \tiny adam & \tiny adam\hspace{1cm}W & \tiny adam\hspace{1cm}W \\
\hline 
\end{tabular}}
\caption{\textbf{\scriptsize{\underline{Optimized Hyper-parameter values per model}}}}
\end{table}\\
In Table 7, below, we explicitly provide the optimised hyper-parameter values for each model on the lexical-based and semantics-based pre-processing methods, respectively. In the case of the SVM models, the obtained values were identical for both pre-processing methods. One can immediately see that some of the obtained optimal hyperparameter values are unusual or uncommon among these models, particularly in the case of the RNN classification models, for both pre-processing methods, respectively, owing to the uniqueness and complexity of our particular dataset for this particular use case and the wide range of tuning values for each hyperparameter.\\
In Table 8, below, we show the model performances of various classification models on the hand-labelled Covid-19 dataset. The results from the VADER and TextBlob algorithms served as a comparison of the degree of similarity or dissimilarity in the criteria used when classifying sentiments via automated means versus classifying sentiments using a manual approach.
\vspace{8cm}
\begin{table}[h]
\centering
\begin{tabular}{|p{0.06\textwidth}|p{0.02\textwidth}|p{0.02\textwidth}|p{0.02\textwidth}|p{0.02\textwidth}|p{0.02\textwidth}|p{0.0225\textwidth}|p{0.02\textwidth}|p{0.02\textwidth}|}
             \hline \multicolumn{9}{|c|}{\tiny{Model Classification Report II}}\\
            \hline
		&  \multicolumn{4}{|c|}{\tiny{NLP-Town BERT}} &  \multicolumn{4}{|c|}{\tiny{Cardiff-NLP RoBERTa}}   \\
		\hline
		\tiny Class & \tiny{Neg} & \tiny{Neu} & \tiny{Pos} & \tiny{All} & \tiny{Neg} & \tiny{Neu} & \tiny{Pos} & \tiny{All} \\
		\hline
		\tiny Precision & \tiny 48  & \tiny 43  & \tiny 46  &  \tiny 46 & \tiny 51  & \tiny 42 & \tiny 45 &  \tiny 46  \\
		\hline
		\tiny Recall & \tiny 43 & \tiny 46 & \tiny 49 &  \tiny 46 & \tiny 47  & \tiny  49  & \tiny 52   &  \tiny 48 \\
		\hline
		\tiny F1-score & \tiny 46 & \tiny 45  & \tiny 48 &  \tiny 46  & \tiny 49 & \tiny 45 & \tiny 48 &  \tiny 47\\
		\hline
		\tiny Accuracy & \multicolumn{3}{c|}{\tiny{}} & \tiny{46} & \multicolumn{3}{|c|}{\tiny{}} & \tiny{47}\\
		\hline
            \multicolumn{9}{|l|}{\tiny{Testing performance of pre-trained BERT and RoBERTa models}}\\
		\hline
		&  \multicolumn{4}{|c|}{\tiny{VADER Algorithm}} &  \multicolumn{4}{|c|}{\tiny{TextBlob Algorithm}}   \\
        \hline
	\tiny Class & \tiny{Neg} & \tiny{Neu} & \tiny{Pos} & \tiny{All} & \tiny{Neg} 
        & \tiny{Neu} & \tiny{Pos} & \tiny{All} \\
		\hline
		\tiny Precision & \tiny49   & \tiny 43  & \tiny 40  &  \tiny 44  & \tiny 44  & \tiny 38 & \tiny 34   &  \tiny 38  \\
		\hline
		\tiny Recall & \tiny 43  & \tiny 31  & \tiny 57  &  \tiny  43 & \tiny 20  & \tiny 37   & \tiny 54    &  \tiny 37  \\
		\hline
		\tiny F1-score & \tiny 46 & \tiny 36 & \tiny 47   &  \tiny 43 & \tiny 28 & \tiny 37 & \tiny 42  & \tiny 36  \\
		\hline
		\tiny Accuracy & \multicolumn{3}{c|}{\tiny{}} & \tiny{43} & \multicolumn{3}{|c|}{\tiny{}} & \tiny{37}\\
		\hline
             \multicolumn{9}{|l|}{\tiny{Testing for  correlation between our Manual and automated labelling}}\\
            \hline
		&  \multicolumn{4}{|c|}{\tiny{Original BERT}} &  \multicolumn{4}{|c|}{\tiny{Original RoBERTa}}   \\
		\hline
		\tiny Class & \tiny{Neg} & \tiny{Neu} & \tiny{Pos} & \tiny{All} & \tiny{Neg} & \tiny{Neu} & \tiny{Pos} & \tiny{All} \\
		\hline
		\tiny Precision & \tiny 45   & \tiny 54   & \tiny 49   &  \tiny 50 & \tiny 49   & \tiny 51  & \tiny 46   &  \tiny 49  \\
		\hline
		\tiny Recall & \tiny 51   & \tiny 46   & \tiny 48 &  \tiny 48  & \tiny 54 & \tiny 47    & \tiny 50 &  \tiny 50 \\
		\hline
		\tiny F1-score & \tiny 48 & \tiny 50  & \tiny 49  &  \tiny 50 & \tiny 51 & \tiny 49  & \tiny 48 & \tiny 49 \\
		\hline
		\tiny Accuracy & \multicolumn{3}{c|}{\tiny{}} & \tiny{48} & \multicolumn{3}{|c|}{\tiny{}} & \tiny{50}\\
		\hline
        \multicolumn{9}{|l|}{\tiny{Testing performance of the Models before fine-tuning on COVID data}}\\
            \hline
		&  \multicolumn{4}{|c|}{\tiny{COVID-19 BERT}} &  \multicolumn{4}{|c|}{\tiny{COVID-19 RoBERTa}}   \\
            \hline
		\tiny Class & \tiny{Neg} & \tiny{Neu} & \tiny{Pos} & \tiny{All} & \tiny{Neg} & \tiny{Neu} & \tiny{Pos} & \tiny{All} \\
		\hline
		\tiny Precision & \tiny 50   & \tiny 57  & \tiny 64  &  \tiny 60  & \tiny  61  & \tiny 58  & \tiny 66   &  \tiny 62  \\
		\hline
		\tiny Recall & \tiny  64  & \tiny 51  & \tiny 69  &  \tiny 61  & \tiny 65  & \tiny  52  & \tiny  58  &  \tiny 61  \\
		\hline
		\tiny F1-score & \tiny 62 & \tiny 54 & \tiny 66 &  \tiny 60  & \tiny 63 & \tiny 55  & \tiny 67 & \tiny 61  \\
		\hline
		\tiny Accuracy & \multicolumn{3}{c|}{\tiny{}} & \tiny{61} & \multicolumn{3}{|c|}{\tiny{}} & \tiny{61}\\
		\hline
          \multicolumn{9}{|l|}{\tiny{Testing  performance of Our Fine-Tuned COVID Models post-training}}\\
            \hline
	\end{tabular}
\caption{\underline{\scriptsize{\textbf{Model Performances on COVID-19 Dataset}}}}
\end{table}
\\In this case, these algorithms showed a poor correlation with the hand labels i.e, overall F1-scores of 43$\%$ and 37$\%$, respectively - which again highlights the superiority of hand-labelling over manual labelling. The pre-trained NLP-Town BERT model when tested on the COVID-19 dataset achieved an overall precision of 46$\%$, an overall accuracy of 46$\%$ yielding an F1-score of 46$\%$, while the pre-trained Cardiff-NLP RoBERTa  model achieved a similar result with an overall precision of 46$\%$, overall accuracy of 48$\%$ yielding an F1-score of 47$\%$. The Original BERT-BASE-CASED model when tested on the COVID-19 dataset achieved an overall precision of 50$\%$, an overall accuracy of 48$\%$ yielding an F1-score of 50$\%$, while our COVID-19 BERT-BASE-CASED  model achieved a much better result with an overall precision of 60$\%$, overall accuracy of 61$\%$ yielding an F1-score of 60$\%$. By comparison, The Original RoBERTa-BASE model when tested on the COVID-19 dataset achieved an overall precision of 49$\%$, an overall accuracy of 50$\%$ yielding an F1-score of 49$\%$, while the COVID-19 RoBERTa-BASE  model achieved a much better result with an overall precision of 62$\%$, overall accuracy of 61$\%$ yielding an F1-score of 61$\%$. The superior performances of the COVID-19 models, when compared to the pre-trained NLP-Town BERT and NLPTown RoBERTa models illustrates both the complexity of the dataset and use-case when compared to sentiment analysis done on much simpler use-cases and using the same labels, as well as the cultural and linguistic differences in the way people communicate in South Africa as compared to the rest of the world. The superior performances of the COVID-19 models, when compared to the Original BERT and RoBERTa models shows that significant training has been achieved.\\
\section*{Acknowledgments}
We give special thanks to the IBM team with whom we had enormous discussion, as well as Malipalema Khang and Abhaya Kumar Swain for technical support during the initial phase of this project.  We also thank Mahnaz Alavinejad for useful discussion. We give a big thank you to Canada’s International Development Research Centre (IDRC) and the Swedish Inter- national Development Cooperation Agency (SIDA) (Grant No. 109559-001) for funding this research. 

\section*{}
\begin{wrapfigure}{l}{3.1cm}
       \centering
    \includegraphics[width=3cm]{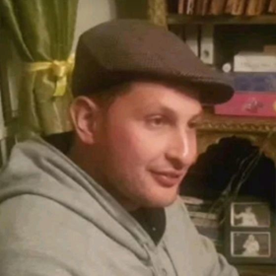}
    \end{wrapfigure}
\textbf{Nicholas Perikli} is an Astrophysics and Biological Sciences graduate with an Honours in Physics currently doing my MSc in Particle Physics at the University of the Witwatersrand. I have many interests in varying fields of academia, enjoy tutoring Maths and Physics, have a lot of experience with Python in general, am well-versed in the field of NLP and privileged to have taken shifts in the ATLAS Control Room during Run 3. I am creative, a fast learner and a deep thinker with an innate and deep curiosity which forces me to question the nature of everything and compels me to continuously learn new things and develop new skills, gain and expand on existing knowledge to develop deeper insights into things and then apply this knowledge and these skills to various research topics and fields. (ORCID ID: 0000-0002-8963-4290).
\section*{}
\vspace{-1.2cm}
\begin{wrapfigure}{l}{3.1cm}
       \centering
        \includegraphics[width=3cm]{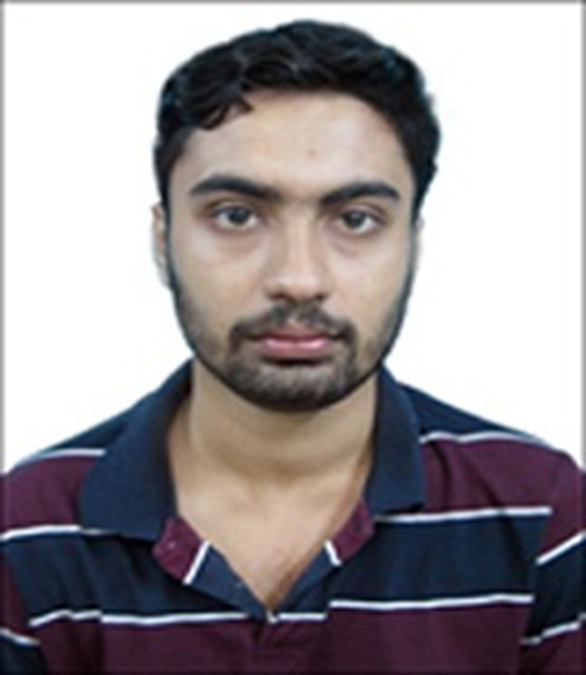}
    \end{wrapfigure}
     \textbf{Srimoy Bhattacharya} has received a Ph.D. in High Energy Physics Phenomenology from the Indian Institute of Technology, Guwahati, (IITG) in 2018. He is currently, a Post-Doctoral Researcher at the University of the Witwatersrand, Johannesburg, South Africa. In addition to the cutting-edge fields in High Energy Physics phenomenology, his current research interests include social media computing, NLP, machine learning for health, and data science for social benefit. Contact Srimoy at bhattacharyasrimoy@gmail.com  (ORCID ID: 0000-0002-9468-5113).\\
\section*{}
\vspace{-1.2cm}
 \begin{wrapfigure}{l}{3.1cm}
       \centering
        \includegraphics[width=3cm]{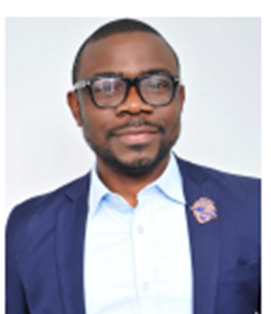}
    \end{wrapfigure}
   \textbf{Blessing Ogbuokiri} received the Ph.D. degree in computer science from the University of the Witwatersrand, Johannesburg, South Africa. He is currently a postdoctoral researcher in the Africa Canada Artificial Intelligence and Data Innovation Consortium (ACADIC) laboratory, Department of Mathematics and Statistics, York University, Toronto, Canada. He has received several academic awards and research grants. He is the recipient of the Dahdaleh Institute Seed Grant for Critical Social Science Perspectives in Global Health Research. His recent research interests include machine learning for health, natural language processing, data science for social good and social media computing. Contact Blessing at blessogb@yorku.ca (ORCID ID:0000-0003-1606-0019). Visit Blessing at (https://www.blessingogbuokiri.com/).\\
\section*{}    
\begin{wrapfigure}{l}{3.1cm}
       \centering
        \includegraphics[width=3cm]{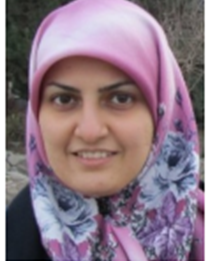}
    \end{wrapfigure}
    \textbf{Zahra Movahedi Nia} is a postdoctoral researcher in the Africa Canada Artificial Intelligence and Data Innovation Consortium (ACADIC) laboratory, York University (Toronto, Canada). She received her PhD in computer engineering, from University of Isfahan (Iran). Her research interests include machine learning, deep learning, and data analytics. Contact Zahra at mova.zhr66@gmail.com (ORCID ID:0000-0002-5528-638X).\\
\section*{}
\vspace{-1.2cm}
\begin{wrapfigure}{l}{3.1cm}
       \centering
        \includegraphics[width=3cm]{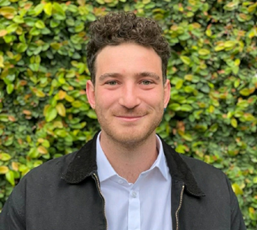}
    \end{wrapfigure}
      \textbf{Benjamin Lieberman} was born in 1994 in Johannesburg, Gauteng province, South Africa. He received his Bachelor of Science (B.Sc)  in the field of nuclear physics and engineering, from the University of Witwatersrand in 2016. He completed his Bachelor of Engineering Honours in Mechanical Engineering at the University of Cape Town in 2018. He began his Master of Science (M.Sc) in the field of Particle Physics at the University of Witwatersrand in collaboration with the ATLAS experiment at the European Organization for Nuclear Research (CERN) in 2019. In 2021 his research was upgraded to a Doctor of Philosophy (PhD) with focus on machine learning applications in particle physics. His research interests lie in the area of applications of semi-supervised machine learning techniques in discovering new bosons at the Large Hadron Collider (LHC), as well as applying scientific modelling and machine learning techniques for epidemiological response. He has specific interest in applying data driven solutions, including machine learning and epidemiological modelling to help inform the COVID-19 response in South Africa. Contact Benjamin at benjamin.lieberman@cern.ch (ORCID ID: 0000-0001-5281-8937).\\
\section*{}
\vspace{-1.2cm}
\begin{wrapfigure}{l}{3.1cm}
       \centering
        \includegraphics[width=3cm]{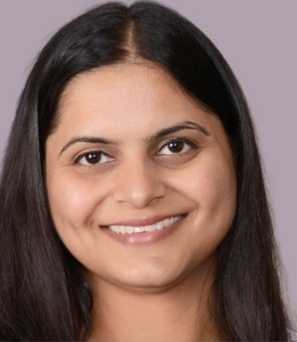}
    \end{wrapfigure}
    \textbf{Nidhi Tripathi} received the master’s degree in computer engineering from Iowa State University, USA in 2017 and is currently a Ph.D. candidate at University of Witwatersrand, South Africa and a member of iThemba labs, South Africa. Her research interest is machine learning based data analysis and modelling. Contact Nidhi at nidhi.tripathi@cern.ch (ORCID ID: 0000-0001-7518-1238). \\
\section*{}  
\vspace{-1.2cm}
\begin{wrapfigure}{l}{3.1cm}
       \centering
        \includegraphics[width=3cm]{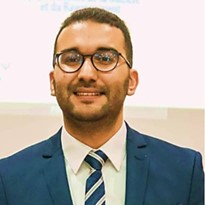}
    \end{wrapfigure}
    \textbf{Dr. Salah-Eddine DAHBI} is presently working as postdoctoral Fellowship in the Institute for Collider Particle Physics at the University of the Witwatersrand and member the ATLAS collaboration at CERN-Large Hadron Collider. Salah’s dissertation research focuses on two topics. The first one addresses the fundamental question of the electroweak symmetry breaking and searching for a new physics predicted in different theoretical models by performing a resonant search. The second part of his PhD thesis is related to the performance of the ATLAS detector at High Luminosity LHC, when the luminosity is expected to be higher, up to ten times the nominal one, and leads to a degradation of the ATLAS sub-components, especially at the forward region closest to the proton-proton interaction point. Dr. Salah-eddine has been conducting a novel project at the Institute for Collider Particle Physics, to search for new resonances beyond the Standard Model of particle physics using machine learning techniques. The creativity in this study is that the method he developed helped the team to reveal the new physics under huge amount of background in the context of weak prior knowledge. Email: salah-eddine.dahbi@cern.ch, (Orchid ID: 0000-0002-5222-7894).\\
\section*{}    
\vspace{-1.2cm}
\begin{wrapfigure}{l}{3.1cm}
       \centering
        \includegraphics[width=3cm]{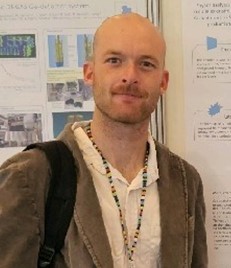}
    \end{wrapfigure}
       \textbf{Finn Stevenson} is an Electro-Mechanical Engineering graduate from the University of Cape Town, currently pursuing a Masters degree through the University of the Witwatersrand in artificial intelligence and data science. My work for the past year has been primarily focused on Covid-19 data modelling in South Africa as a member of the Wits-iThemba Covid-19 Modelling team, the official modelling team of the Gauteng province. We have provided valuable AI-driven tools related to Covid-19 case data predictions, economic recovery predictions, data-informed policy recommendations and the development of early alert systems.  During the period of working on Covid-19 related content, I have learnt many skills and connected with many interesting people due to the urgency and necessity of the work. I am now working on applying techniques I have learnt to data from the LHC at CERN for anomaly detection in Beyond the Standard Model searches as part of the ATLAS experiment. In the future, I hope to apply the broad skill set I have developed to many exciting projects. I am most passionate about work that is novel and experimental in nature. Outside of the office (home office), I am a keen adventurer and love being outdoors, on the mountain, in the forest or in the ocean. I am an avid surfer, skateboarder and hiker. I never miss the opportunity of a spontaneous adventure whatever that might entail. Contact Finn at finn.stevenson@cern.ch (ORCID ID:0000-0003-0444-2992).\\
\section*{}    
\vspace{+1.2cm}
\begin{wrapfigure}{l}{3.1cm}
       \centering
        \includegraphics[width=3cm]{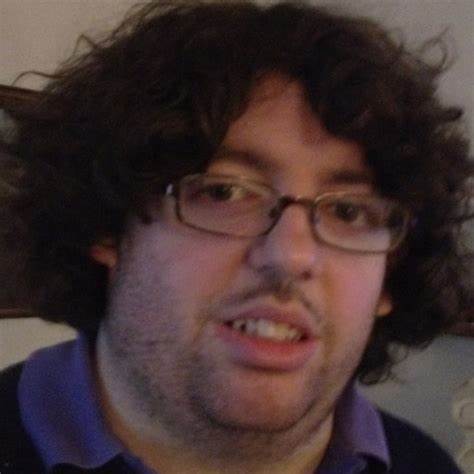}
    \end{wrapfigure}
    \textbf{Nicola Luigi Bragazzi} got his MD in general medicine and surgery from Genoa University (Genoa, Italy) in 2011, his PhD in biophysics from Marburg University (Marburg, Germany) in 2014 and his specialization in Public Health from Genoa University (Genoa, Italy) in 2017. He is currently with the Department of Food and Drugs, University of Parma, Parma, Italy. Contact Bragazzi at robertobragazzi@gmail.com (ORCID ID: (0000-0001-8409-868X).\\
\section*{} 
\vspace{-1.2cm}
\begin{wrapfigure}{l}{3.1cm}
       \centering
        \includegraphics[width=3cm]{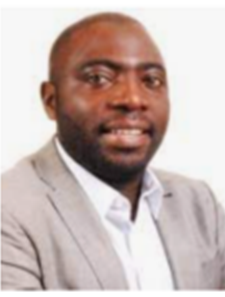}
    \end{wrapfigure}
      \textbf{Jude Kong} is the Director of the Africa-Canada Artificial Intelligence and Data Innovation Consortium (ACADIC). He leads two other networks of researchers that are designing early warning frameworks for emerging infectious disease outbreaks: Mathematics for Public Health (MfPH) and One Health Modelling Network for Emerging Infectious Diseases (OMNI). He is an expert in mathematical modelling, artificial intelligence, data science, infectious disease modelling and mathematics education. His principal research program focuses on the use of quantitative methods to improve decision-making for epidemic and pandemic prevention, preparedness and response. In 2020, he won a York Research Leader Award. In 2021 he was spotlighted among Canadian Innovation Research Leaders 2021(https://researchinfosource.com/pdf/CIL2021\hspace{1cm}.pdf) for his work with ACADIC. In 2022, he was spotlighted as a Change Maker by People of York University for his work in helping others learn mathematical concepts and encouraging them to find their passion and achieve more than they thought was possible (https://www.yorku.ca/peopleofyu/2022/02/18/ jude-kong-faculty/). Contact Jude Kong at jdkong@yorku.ca (ORCID ID: 0000-0002-7557-5672).\\
 \section*{}   
 \begin{wrapfigure}{l}{3.1cm}
       \centering
        \includegraphics[width=3cm]{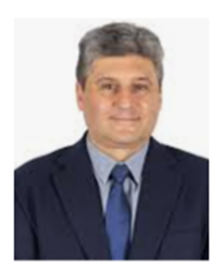}
    \end{wrapfigure}
   Prof.\textbf{ Bruce Mellado} is the Co-president of Africa Canada Artificial Intelligence and Data Innovation Consortium (ACADIC) and a member of the Gauteng Premier’s COVID-19 Advisory Committee, where he leads work on predicting and forecasting the dynamics of COVID-19. He is the recipient of several awards and fellowships. He is an Internationally acclaimed, B1-rated researcher of the National Research Foundation of South Africa. He is an expert in Artificial Intelligence. Contact Bruce at bruce.mellado.garcia@cern.ch (ORCID ID: 0000-0003-4838-1546).\\
\end{document}